\documentclass[aps,pre,onecolumn,floatfix]{revtex4}

\usepackage{graphicx}
\usepackage{amssymb,amsfonts,amsmath}
\usepackage{epsf}
\usepackage{subfigure}
\usepackage{epstopdf}
\usepackage{color}
\usepackage{ulem}
\usepackage{mathrsfs}

\newcommand{\be}{\begin{equation}}
\newcommand{\ee}{\end{equation}}
\newcommand{\bea}{\begin{eqnarray}}
\newcommand{\eea}{\end{eqnarray}}

\begin{document}

\title{Finite-time fluctuation theorem for diffusion-influenced surface reactions on \\ spherical and Janus catalytic particles}

\author{Pierre Gaspard}
\email{gaspard@ulb.ac.be}
\affiliation{ Center for Nonlinear Phenomena and Complex Systems, Universit{\'e} Libre de Bruxelles (U.L.B.), Code Postal 231, Campus Plaine, B-1050 Brussels, Belgium}

\author{Patrick Grosfils}
\email{Patrick.Grosfils@ulb.ac.be}
\affiliation{ Center for Nonlinear Phenomena and Complex Systems, Universit{\'e} Libre de Bruxelles (U.L.B.), Code Postal 231, Campus Plaine, B-1050 Brussels, Belgium}

\author{Mu-Jie Huang}
\email{mjhuang@chem.utoronto.ca}
\affiliation{ Chemical Physics Theory Group, Department of Chemistry, University of Toronto, Toronto, Ontario M5S 3H6, Canada}

\author{Raymond Kapral}
\email{rkapral@chem.utoronto.ca}
\affiliation{ Chemical Physics Theory Group, Department of Chemistry, University of Toronto, Toronto, Ontario M5S 3H6, Canada}

\begin{abstract}
A finite-time fluctuation theorem for the diffusion-influenced surface reaction ${\rm A}\rightleftharpoons{\rm B}$ is investigated for spherical and Janus catalytic particles.  The finite-time rates and thermodynamic force are analytically calculated by solving diffusion equations with the special boundary conditions of the finite-time fluctuation theorem.  Theory is compared with numerical simulations carried out with two different methods: a random walk algorithm and multiparticle collision dynamics.
\end{abstract}

\maketitle

\section{Introduction}

Away from equilibrium, currents of energy and matter flow across open systems.  Because of the atomic structure of matter, these currents manifest fluctuations, which can be characterized by their large-deviation properties within the framework of probability theory.  Furthermore, the motions of microscopic particles obey fundamental time-reversal symmetry, which implies that the current fluctuations satisfy the so-called fluctuation theorems \cite{ECM93,G96,K98,LS99,J11,G04JCP,AG04,AG06,DDR04,D07,SIPGD07,PGPD10,PGLD13,G13NJP,BDGJL15}.  These fluctuation theorems are usually obtained in the long-time limit.  Remarkably, fluctuation theorems may also hold at every finite time under specific conditions.  This is the case for Markov jump processes describing linear reactions in homogeneous systems \cite{AG08}.  Recently, a finite-time fluctuation theorem was established for diffusion-influenced surface reactions in spatially extended systems described by stochastic partial differential equations \cite{TDFT}.

Fluctuation theorems play a key role in the determination of the thermodynamic forces driving systems into nonequilibrium steady states.  Under such circumstances, the measurement of the driving forces is carried out in the long-time limit.  In systems where a finite-time fluctuation theorem holds, the driving forces are defined at every finite time, and this allows one to investigate how they vary in time and to determine the time scale on which they  converge to their asymptotic values.

The purpose of this paper is to investigate aspects of the finite-time fluctuation theorem for the diffusion-influenced surface reaction ${\rm A}\rightleftharpoons{\rm B}$ on spherical and Janus catalytic particles.  We deduce analytical expressions for the time dependence of the thermodynamic force of the reactive process by solving macroscopic diffusion equations with the special boundary conditions obtained in Ref.~\cite{TDFT}. This solution provides the large-deviation properties of the random number of reactive events occurring during finite-time intervals.  The theoretical results are compared with numerical simulations carried out using two different methods.  In the first method, the process is simulated with particles carrying a color (A or B) and independently diffusing according to a random walk algorithm between the reservoir and the catalytic surface where they may interchange their color.  In the second method, the simulation is performed using the multiparticle collision dynamics \cite{K08}.  In this method, the system comprises reactive and solvent particles, together with a spherical composite catalytic particle made from a collection of linked beads \cite{HSGK18}, some of which catalyze the reaction ${\rm A}\rightleftharpoons{\rm B}$.  Since the catalytic particle is larger than the reacting and solvent particles, its motion is small and negligible, allowing us to make comparisons with the theory that assumes immobile catalytic surfaces.

The paper is organized as follows.  The principal results of the finite-time fluctuation theorem are summarized in section~\ref{Sec-thm}.  The theory is applied to a spherical catalytic particle in section~\ref{S-geom}, and a Janus catalytic particle in section~\ref{J-geom}.  In these sections, the deterministic diffusion equations are solved for the geometries of the particles in these systems.  Complete analytic expressions are obtained for the time dependence of the rates and the corresponding affinity.  For each system, theory is compared with numerical simulations.  Section~\ref{conclusion} gives concluding remarks and perspectives.


\section{The finite-time fluctuation theorem}
\label{Sec-thm}

We consider a system where the molecular species A and B diffuse in a three-dimensional domain $V$ extending between three surfaces $\partial V = S_{\rm cat}\cup S_{\rm inert}\cup S_{\rm res}$.  The reaction ${\rm A}\rightleftharpoons{\rm B}$ takes place at the catalytic surface $S_{\rm cat}$.  The molecules ${\rm A}$ and ${\rm B}$ are reflected at the inert surface $S_{\rm inert}$.  Moreover, they enter and exit the domain $V$ at the surface $S_{\rm res}$ in contact with a reservoir.  Accordingly, the concentrations (i.e., the densities) $c_k$ of the species $k\in\{{\rm A},{\rm B}\}$ are ruled by the fluctuating diffusion equations
\be
\partial_t \, c_k + \pmb{\nabla}\cdot {\bf j}_k = 0 \, ,\qquad\mbox{with}\qquad  {\bf j}_k  = -D_k \pmb{\nabla} c_k + \pmb{\eta}_k\, , \label{diff-eq-k}
\ee
and boundary conditions
\bea
{\rm if}\ {\bf r}\in S_{\rm cat} : \qquad  && D_{\rm A} \, \partial_{\bot} c_{\rm A}({\bf r},t) = -D_{\rm B}\, \partial_{\bot}c_{\rm B}({\bf r},t) =\kappa_+ c_{\rm A}({\bf r},t) - \kappa_- c_{\rm B}({\bf r},t)+ \xi({\bf r},t) \, , \label{bc-cat}\\
{\rm if}\ {\bf r}\in S_{\rm inert} : \qquad && \partial_{\bot} c_k({\bf r},t) = 0 \, , \label{bc-inert}\\
{\rm if}\ {\bf r}\in S_{\rm res} : \qquad && c_k({\bf r},t) = \bar{c}_k \, ,  \label{bc-res}
\eea
where  $D_k$ are the diffusion coefficients, $\partial_{\bot}$ is the gradient in the direction normal to the surface and oriented towards the interior of the domain~$V$, $\kappa_{\pm}$ are the surface reaction rate constants, and $\bar{c}_k$ the concentration values at the reservoir. The Gaussian noises associated with bulk diffusion are characterized by
\be
\langle\pmb{\eta}_k({\bf r},t)\rangle = 0 \, , \qquad \langle\pmb{\eta}_k({\bf r},t)\otimes\pmb{\eta}_{k'}({\bf r'},t')\rangle = 2\,  D_k \, c_k({\bf r}, t) \, \delta_{kk'} \, \delta({\bf r}-{\bf r'}) \, \delta(t-t') \, {\boldsymbol{\mathsf 1}} \, ,
\ee
where $k,k'\in\{{\rm A},{\rm B}\}$ and ${\boldsymbol{\mathsf 1}}$ is the $3\times 3$ identity matrix, while the Gaussian noise associated with surface reaction satisfies
\be
\langle \xi({\bf r},t) \rangle = 0 \, , \qquad
\delta^{\rm s}({\bf r})\, \langle \xi({\bf r},t)\,\xi({\bf r'},t') \rangle\, \delta^{\rm s}({\bf r'}) = (\kappa_+\, c_{\rm A}+\kappa_-\, c_{\rm B}) \, \delta^{\rm s}({\bf r}) \, \delta({\bf r}-{\bf r'}) \, \delta(t-t')  \, ,
\ee
which is expressed in terms of surface delta distributions $\delta^{\rm s}({\bf r})$ nonvanishing if ${\bf r}\in S_{\rm cat}$ \cite{BAM76}.

As proved in Ref.~\cite{TDFT}, the probability $P(n,t)$ that the net number $n$ of reactive events ${\rm A}\to{\rm B}$ have occurred during the time interval $[0,t]$ satisfies the finite-time fluctuation theorem
\be
\boxed{\frac{P(n,t)}{P(-n,t)} = \exp ( {\cal A}_t \, n )}
\label{FT}
\ee
{\it at every time}.  The time-dependent affinity is defined as
\be
{\cal A}_t = \ln\frac{W_t^{(+)}}{W_t^{(-)}}
\label{A-dfn}
\ee
with the finite-time rates
\be
W_t^{(\pm)} = W_{\infty}^{(\pm)} + \frac{1}{t}\, \Psi(t) \, , \label{W}
\ee
having the asymptotic values
\be
W_{\infty}^{(+)} = \Sigma \kappa_+ \bar{c}_{\rm A} \,  , \qquad\qquad
W_{\infty}^{(-)} = \Sigma \kappa_- \bar{c}_{\rm B} \, , \label{W-inf}
\ee
and a time dependence given by the function
\be
\Psi(t) = \ell^2 \kappa_+\kappa_- \left[ \frac{\bar{c}_{\rm B}}{D_{\rm A}^2} \Upsilon_{\rm A}(t) + \frac{\bar{c}_{\rm A}}{D_{\rm B}^2} \Upsilon_{\rm B}(t)\right] .
\label{Psi}
\ee
The asymptotic values~(\ref{W-inf}) are expressed in terms of the effective catalytic surface area
\be
\Sigma = \int_{\rm cat} dS \, (1-\phi) \, ,
\label{Sigma}
\ee
where $\phi$ is the solution of the Laplace equation,
\be
\nabla^2 \phi = 0 \, , \label{phi1}
\ee
with the boundary conditions
\be
\left(\partial_{\bot} \phi\right)_{\rm cat} = \ell^{-1}(\phi-1)_{\rm cat} \, , \qquad \left(\partial_{\bot} \phi\right)_{\rm inert} = 0 \, , \qquad\mbox{and}\qquad \left(\phi\right)_{\rm res} = 0 \, , \label{phi4}
\ee
involving the characteristic length of the diffusion-influenced surface reaction,
\be
\ell \equiv \left(\frac{\kappa_+}{D_{\rm A}} + \frac{\kappa_-}{D_{\rm B}}\right)^{-1} .
\label{ell}
\ee
The stationary concentration fields are thus given for $k={\rm A},{\rm B}$ by
\be
\langle c_k\rangle_{\rm st} = \bar{c}_k +\frac{\nu_k\ell}{D_k}   \left(\kappa_+\bar{c}_{\rm A}-\kappa_- \bar{c}_{\rm B}\right)  \phi \, ,
\label{NESS}
\ee
where $\nu_{\rm A}=-1$ and $\nu_{\rm B}=+1$ are the stoichiometric coefficients of the surface reaction.

Equation~(\ref{Psi}) is written in terms of the functions
\be
\Upsilon_k(t) = \int dV \phi({\bf r})\left[ \phi({\bf r})-f_k({\bf r},t)\right] \, ,
\label{Upsilon}
\ee
where $f_k$ is the solution of the diffusion equations,
\be
\partial_t f_k = D_k \nabla^2 f_k \, , \label{f1}
\ee
with the boundary and initial conditions
\be
\left(\partial_{\bot} f_k\right)_{\rm cat} = \left(\frac{\kappa_+}{D_{\rm A}}\, f_{\rm A}+\frac{\kappa_-}{D_{\rm B}}\, f_{\rm B}\right)_{\rm cat} \, , \qquad \left(\partial_{\bot} f_k\right)_{\rm inert} = 0 \, , \qquad
\left(f_k\right)_{\rm res} = 0 \, , \qquad \left(f_k\right)_{t=0} = \phi \, , \label{f5}
\ee
for $k={\rm A},{\rm B}$.

\begin{figure}[h]
\centerline{\scalebox{0.42}{\includegraphics{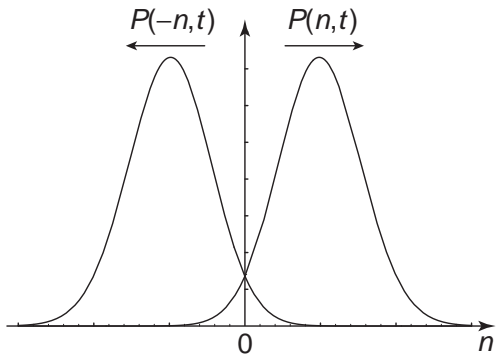}}}
\caption{Schematic representation of the probability distributions $P(\pm n,t)$ of opposite fluctuations in the number $n$ of reactive events occurring during the time interval $[0,t]$ under nonequilibrium conditions.  These distributions shift away from the origin $n=0$ as time $t$ increases, so that their overlap rapidly decreases.}
\label{fig1}
\end{figure}

The mean current, i.e., the mean overall reaction rate, is given by
\be
{\cal J}  = W_t^{(+)} - W_t^{(-)} = \Sigma\left( \kappa_{+}\bar{c}_{\rm A} - \kappa_{-}\bar{c}_{\rm B} \right) \, , \label{J}
\ee
which does not depend on time because of the expression~(\ref{W}) for the rates; however, the diffusivity of the current fluctuations
\be
{\cal D}_t  = \frac{1}{2} \left(W_t^{(+)} +W_t^{(-)}\right) \label{D_t}
\ee
does depend on time.

We note that the rates~(\ref{W}) are determined at early times by the stationary solutions~(\ref{NESS}) according to
\bea
&& W_t^{(+)} = \int_{\rm cat} dS \, \kappa_+\langle c_{\rm A}\rangle_{\rm st} + O(t) \, , \label{W+early}\\
&& W_t^{(+)} = \int_{\rm cat} dS \, \kappa_- \langle c_{\rm B}\rangle_{\rm st} +O(t) \, . \label{W-early}
\eea

As schematically depicted in figure~\ref{fig1}, the probability distributions $P(\pm n,t)$ shift in opposite directions under nonequilibrium conditions, because their mean values are given by $\langle n\rangle_t=\pm{\cal J}t$ in terms of the mean current~(\ref{J}).  If the concentrations at the reservoir satisfy the equilibrium condition $\kappa_{+}\bar{c}_{\rm A} = \kappa_{-}\bar{c}_{\rm B}$, the time-dependent rates~(\ref{W}) are equal and the mean current~(\ref{J}) vanishes together with the affinity~(\ref{A-dfn}).  Detailed balance is thus recovered at equilibrium.

In the following sections, these results, proved in Ref.~\cite{TDFT}, are applied to spherical and Janus catalytic particles.


\section{Spherical catalytic particle}
\label{S-geom}

\subsection{Theory}
\label{S-geom-theory}

We consider a spherical catalyst of radius $r=R$ centered on the origin of the reference frame, and a reservoir for species A and B that is located at a sphere of radius $r=L\gg R$, $r$ denoting the radial distance from the origin.  For this spherical geometry, the Laplacian operator in the problem~(\ref{phi1})-(\ref{phi4}) for the stationary solution reduces to
\be
\nabla^2\phi =\frac{1}{r}\frac{d^2}{dr^2}\left(r\, \phi \right) = 0 \, ,
\ee
while $\partial_{\bot}=d/dr$.  Hence, the equations can be solved by setting $\phi(r)=u(r)/r$, yielding the solution
\be
\phi(r) = R\, \frac{\rm Da}{\tilde\Delta} \left(\frac{1}{r}-\frac{1}{L}\right) ,
\label{phi}
\ee
where
\be
{\rm Da}\equiv \frac{R}{\ell} = R \left(\frac{\kappa_+}{D_{\rm A}} + \frac{\kappa_-}{D_{\rm B}}\right)
\label{Damkohler}
\ee
is the Damk\"ohler number defined in terms of the characteristic length~(\ref{ell}), and
\be
\tilde\Delta \equiv 1 + {\rm Da} \left(1 - \frac{R}{L}\right) .
\label{tilde-Delta-3D}
\ee
The effective catalytic surface area~(\ref{Sigma}) is given by $\Sigma = 4\pi R^2/\tilde\Delta = 4\pi R^2/[1 + {\rm Da}\,(1 - R/L)]$.
Setting $f_k(r,t)=v_k(r,t)/r$, the functions~(\ref{Upsilon}) can be expressed as
\be \label{up}
\Upsilon_k(t) =4\pi \int_R^L dr \, u(r) \left[ u(r)-v_k(r,t)\right] .
\ee
The solutions of equations~(\ref{f1})-(\ref{f5}) can be expanded as
\be
v_k(r,t) = \sum_{i=1}^{\infty} a_i \, {\rm e}^{-D_k q_i^2 t} \sin q_i(L-r) + \chi_k \sum_{i=1}^{\infty} \tilde a_i \, {\rm e}^{-D_k \tilde q_i^2 t} \sin \tilde q_i(L-r) \label{vk-3D}
\ee
with the coefficients $\chi_{\rm A} = D_{\rm A}/\kappa_+$ and $\chi_{\rm B} = -D_{\rm B}/\kappa_-$
and where the eigenvalues are the roots of
\bea
&& q_i R = -\Delta \, \tan q_i (L-R) \, , \label{eigen-3D}\\
&& \tilde q_i R = -\tan \tilde q_i (L-R) \, ,
\eea
with $\Delta \equiv 1 + {\rm Da} = \lim_{L\to\infty}\tilde\Delta$ and $q_i,\tilde q_i>0$.  Since the solutions~(\ref{vk-3D}) satisfy the initial conditions $v_k(r,t=0)=u(r)$ for $k={\rm A}\,,{\rm B}$, we find that the expansion coefficients are given by
\be
a_i = \frac{4 R^2\left[ \sin q_i (L-R) - q_i(L-R) \cos q_i (L-R)\right]}{q_i L\ell \tilde\Delta \left[2q_i(L-R)-\sin 2q_i (L-R)\right]}
\ee
and $\tilde a_i=0$.  Using equation~(\ref{eigen-3D}), the functions~(\ref{Upsilon}) are obtained as
\be
\Upsilon_k(t) = \frac{8\pi R^2}{\ell^2(L-R)} \sum_{i=1}^{\infty} \frac{1-{\rm e}^{-D_k q_i^2t}}{q_i^2\left[q_i^2+\frac{\Delta\tilde\Delta}{R^2(1-R/L)}\right]} .
\ee

Therefore, the rates are given by equation~(\ref{W}) with
\be
W_{\infty}^{(+)} = \frac{4\pi R^2}{\tilde\Delta} \, \kappa_+ \bar{c}_{\rm A} \, , \qquad\qquad
W_{\infty}^{(-)} = \frac{4\pi R^2}{\tilde\Delta} \, \kappa_- \bar{c}_{\rm B} \,  ,
\label{W-inf-3D}
\ee
and
\be
\Psi(t) = \frac{4\pi R^5\kappa_+\kappa_-}{\left(\Delta\tilde\Delta\right)^{3/2}} \left(1-\frac{R}{L}\right)^{3/2}\left[ \frac{\bar{c}_{\rm B}}{D_{\rm A}^2}\, \Omega_{\rm s}\left(\gamma_{\rm A}t\right) + \frac{\bar{c}_{\rm A}}{D_{\rm B}^2}\, \Omega_{\rm s}\left(\gamma_{\rm B}t\right)\right]
\label{Psi-3D}
\ee
with the function
\be
\Omega_{\rm s}(\tau) \equiv \frac{2R}{L\sqrt{\Delta\tilde\Delta(1-R/L)}} \sum_{i=1}^{\infty} \frac{1-{\rm e}^{-Q_i^2\tau}}{Q_i^2\left(Q_i^2+1\right)}
\label{Omega-s}
\ee
obtained after the substituting $Q_i=q_iR\sqrt{(L-R)/(L\Delta\tilde\Delta)}$ and defining the rates
\be
\gamma_k \equiv \frac{D_k\Delta\tilde\Delta}{R^2\left(1-R/L\right)} \, .
\ee
Using these results, $\Psi(t)$ can be determined by numerical evaluation of the sums.

Using the long-time limit of equations~(\ref{up}) and (\ref{vk-3D}) we find that the function~(\ref{Psi}) is given by
\be
\Psi({\infty}) = \frac{4\pi R^4 L \kappa_+\kappa_-}{3\tilde\Delta^2} \left(1-\frac{R}{L}\right)^3 \left(\frac{\bar{c}_{\rm A}}{D_{\rm B}^2} + \frac{\bar{c}_{\rm B}}{D_{\rm A}^2} \right) .
\label{Psi-3D-t_inf}
\ee
We see that this asymptotic value is proportional to the distance $L$ to the reservoirs and thus diverges in the limit $L\to\infty$.  Comparing equation~(\ref{Psi-3D-t_inf}) with the limit $t\to\infty$ of equation~(\ref{Psi-3D}), we get
\be
\Omega_{\rm s}(\infty) = \frac{L\Delta^{3/2}}{3R\tilde\Delta^{1/2}} \left(1-\frac{R}{L}\right)^{3/2} \, .
\ee
Consequently, on the long time scale $D_k^{-1}L^2 \ll t$ for $k\in\{{\rm A},{\rm B}\}$,
the finite-time affinity behaves as
\be
{\cal A}_t = \ln\frac{\kappa_+\bar{c}_{\rm A} +\frac{1}{t} \frac{R^2L\kappa_+\kappa_-}{3\tilde\Delta}\left(1-\frac{R}{L}\right)^3\left(\frac{\bar{c}_{\rm A}}{D_{\rm B}^2}+\frac{\bar{c}_{\rm B}}{D_{\rm A}^2}\right)+O\left(t^{-1}{\rm e}^{-D_{\rm A}q_1^2t}\right) + O\left(t^{-1}{\rm e}^{-D_{\rm B}q_1^2t}\right)}{\kappa_-\bar{c}_{\rm B} +\frac{1}{t} \frac{R^2L\kappa_+\kappa_-}{3\tilde\Delta}\left(1-\frac{R}{L}\right)^3\left(\frac{\bar{c}_{\rm A}}{D_{\rm B}^2}+\frac{\bar{c}_{\rm B}}{D_{\rm A}^2}\right)+O\left(t^{-1}{\rm e}^{-D_{\rm A}q_1^2t}\right) + O\left(t^{-1}{\rm e}^{-D_{\rm B}q_1^2t}\right)} \, .
\ee

If $L\gg R$, the rates~(\ref{W}) have the well-defined asymptotic values
\be
W_{\infty}^{(+)} = \frac{4\pi R^2}{1+{\rm Da}} \, \kappa_+ \bar{c}_{\rm A} \, , \qquad\qquad
W_{\infty}^{(-)} = \frac{4\pi R^2}{1+{\rm Da}} \,  \kappa_- \bar{c}_{\rm B} \, ,
\label{W-inf-3D-lim}
\ee
and
\be
\Psi(t) = \frac{4\pi R^5\kappa_+\kappa_-}{\left(1+{\rm Da}\right)^3} \left[ \frac{\bar{c}_{\rm B}}{D_{\rm A}^2} \, \Omega\left(\gamma_{\rm A}t\right) + \frac{\bar{c}_{\rm A}}{D_{\rm B}^2}\, \Omega\left(\gamma_{\rm B}t\right)\right] ,
\label{Psi-3D-lim}
\ee
where $\gamma_k = D_kR^{-2}( 1+{\rm Da})^2$
and the function $\Omega_{\rm s}(\tau)$ is given by its integral approximation,
\be
\Omega(\tau) = \frac{1}{\pi} \int_{-\infty}^{+\infty} \frac{1-{\rm e}^{-Q^2 \tau}}{Q^2 \left(1 + Q^2\right)}\, dQ = 2\sqrt{\frac{\tau}{\pi}} -1 + {\rm e}^{\tau}\, {\rm erfc}(\sqrt{\tau})\, .
\label{Omega}
\ee
This function has the following asymptotic expansion for $\tau\to\infty$
\be
\Omega(\tau) = 2\sqrt{\frac{\tau}{\pi}} -1 + \frac{1}{\sqrt{\pi\tau}} + O\left(\frac{1}{\tau^{3/2}}\right) \, ,
\label{Omega-asympt}
\ee
and Taylor series around $\tau=0$
\be
\Omega(\tau) = \tau - \frac{4}{3\sqrt{\pi}}\, \tau^{3/2} + \frac{1}{2}\, \tau^2 - \frac{8}{15\sqrt{\pi}}\, \tau^{5/2} +O(\tau^3) \, .
\label{Omega-series}
\ee
Consequently, the function~(\ref{Psi-3D-lim}) increases without limit as $\sqrt{t}$.  However, it is divided by the time $t$ in the expression~(\ref{W}) for the rates, which thus have the well-defined values~(\ref{W-inf-3D-lim}) in the long-time limit.

On the intermediate time scale $D_k^{-1}R^2/(1+{\rm Da})^2 \ll t \ll D_k^{-1}L^2$, the rates have the following expressions
\be
W_t^{(\pm)} = W_{\infty}^{(\pm)} + \frac{2}{\sqrt{\pi t}} \frac{4\pi R^4 \kappa_+\kappa_-}{(1+{\rm Da})^2} \left(\frac{\bar{c}_{\rm A}}{D_{\rm B}^{3/2}}+\frac{\bar{c}_{\rm B}}{D_{\rm A}^{3/2}}\right) +O(t^{-1})
\label{Q-long-3D}
\ee
in terms of their asymptotic values~(\ref{W-inf-3D-lim}).
Accordingly, the affinity behaves as
\be
{\cal A}_t = \ln\frac{\kappa_+\bar{c}_{\rm A}+\frac{2}{\sqrt{\pi t}}\frac{R^2\kappa_+\kappa_-}{1+{\rm Da}}\left(\frac{\bar{c}_{\rm A}}{D_{\rm B}^{3/2}}+\frac{\bar{c}_{\rm B}}{D_{\rm A}^{3/2}}\right)+O(t^{-1})}{\kappa_-\bar{c}_{\rm B}+\frac{2}{\sqrt{\pi t}}\frac{R^2\kappa_+\kappa_-}{1+{\rm Da}}\left(\frac{\bar{c}_{\rm A}}{D_{\rm B}^{3/2}}+\frac{\bar{c}_{\rm B}}{D_{\rm A}^{3/2}}\right)+O(t^{-1})} .
\label{Aff-inter-3D}
\ee
In the long-time limit, the affinity~(\ref{A-dfn}) thus converges towards the expected asymptotic value
\be
{\cal A}_{\infty} = \ln \frac{\kappa_{+}\bar{c}_{\rm A}}{\kappa_{-}\bar{c}_{\rm B}} .
\label{A-inf}
\ee

On the short time scale $t \ll D_k^{-1}R^2/(1+{\rm Da})^2$,
equations~(\ref{W+early}) and~(\ref{W-early}) with equation~(\ref{NESS}) show that the rates are given at early times by
\bea
W_t^{(+)} &=& \frac{4\pi R^2\kappa_+}{\tilde\Delta} \left[\bar{c}_{\rm A}+\kappa_-R\left(1-\frac{R}{L}\right)\left(\frac{\bar{c}_{\rm A}}{D_{\rm B}}+\frac{\bar{c}_{\rm B}}{D_{\rm A}}\right)\right] + O(t) \, , \nonumber\\
W_t^{(-)} &=&  \frac{4\pi R^2\kappa_-}{\tilde\Delta} \left[\bar{c}_{\rm B}+\kappa_+R\left(1-\frac{R}{L}\right)\left(\frac{\bar{c}_{\rm A}}{D_{\rm B}}+\frac{\bar{c}_{\rm B}}{D_{\rm A}}\right)\right] + O(t)\, ,
\eea
and the corresponding affinity is
\be
{\cal A}_t = \ln\frac{\kappa_+\bar{c}_{\rm A}+\kappa_+\kappa_-R\left(1-\frac{R}{L}\right)\left(\frac{\bar{c}_{\rm A}}{D_{\rm B}}+\frac{\bar{c}_{\rm B}}{D_{\rm A}}\right)+O(t)}{\kappa_-\bar{c}_{\rm B}+\kappa_+\kappa_-R\left(1-\frac{R}{L}\right)\left(\frac{\bar{c}_{\rm A}}{D_{\rm B}}+\frac{\bar{c}_{\rm B}}{D_{\rm A}}\right)+O(t)} \, ,
\ee
which holds for small enough times.  Therefore, the affinity can take an early-time value that is much smaller than its asymptotic value~(\ref{A-inf}).

Taking equation~(\ref{Aff-inter-3D}) with $D=D_{\rm A}=D_{\rm B}$ and $\kappa=\kappa_{\pm}$, we see that the affinity reaches its asymptotic value if $t \gg D^{-1} R^2/(1+{\rm Da}^{-1})^2$.
This condition reads $t \gg D^{-1} R^2$ in the diffusion-limited regime where ${\rm Da}\gg 1$,
and $t \gg {\rm Da}^2 D^{-1} R^2$ in the reaction-limited regime where ${\rm Da}\ll 1$.
Since the molecular diffusivities typically take the value $D\simeq 10^{-9}$ m$^2$/s, the crossover time for a catalytic particle of micrometric radius is of the order of milliseconds.

\subsection{Numerical results}

For the simulation results, we suppose that the diffusion coefficients and the forward and reverse rate constants are equal, $D\equiv D_{\rm A}=D_{\rm B}$ and $\kappa\equiv \kappa_{\pm}$.  The function $\Psi(t)$ in equation~(\ref{Psi}) is given by
\be
\Psi(t) = \frac{\ell^2 \kappa^2}{D^2}  \left( \bar{c}_{\rm A}+ \bar{c}_{\rm B}\right) \, \Upsilon(t)
\label{Psi-3D-equal}
\ee
with
\be
\Upsilon(t) = 4\pi \int_R^L dr \, u(r) \, \left[u(r)-v(r,t)\right] ,
\ee
where $u(r) = \ell^{-1}\tilde\Delta^{-1} R^2 \left(1-r/L\right)$ and $\tilde\Delta \equiv 1 + 2\kappa D^{-1}R\left(1 - R/L\right)$.
The function $v(r,t)$ is the solution of the following problem:
\bea
&&\partial_t v(r,t)= D\,\partial_r^2 v(r,t) \, , \\
&& R\, \partial_r v(R,t) = \left( 1 + \frac{2\kappa R}{D} \right) v(R,t) \, , \\
&& v(L,t) = 0 \, , \\
&& v(r,0)=u(r) \, ,
\eea
which can be solved numerically by spatial discretization into $I$ cells of size $\Delta r= (L-R)/I$.

\begin{figure}[h]
\centerline{\scalebox{0.5}{\includegraphics{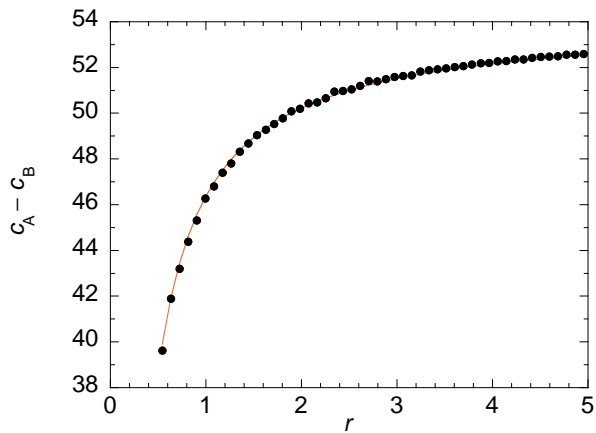}}}
\caption{Spherical catalytic particle: Stationary profile of the concentration difference $c_{\rm A}-c_{\rm B}$ versus the radial distance $r$ for the simulation with the parameter values~(\ref{3D-ex}). The dots are the simulation data and the solid line the theoretical result.}
\label{fig2}
\end{figure}

\begin{figure}[h]
\centerline{\scalebox{0.5}{\includegraphics{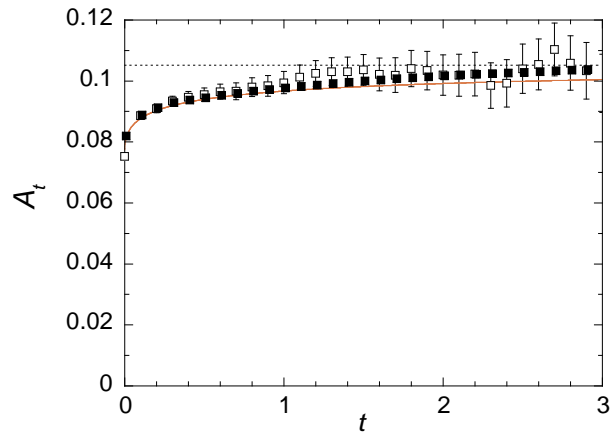}}}
\caption{Spherical catalytic particle: Affinity ${\cal A}_t$ versus time $t$ for the simulation with the parameter values~(\ref{3D-ex}).  The dots are the simulation data and the solid line the theoretical result. The open squares show the affinity directly measured with $\ln[P(n,t)/P(-n,t)]$, while the filled squares show the affinity obtained from the Gaussian fit~(\ref{Gaussian_fit}).}
\label{fig3}
\end{figure}

These theoretical expressions are compared with numerical simulations based on a diffusive random walk model as well as multiparticle collision dynamics.

We first discuss the results obtained when the A and B particles move according to a diffusive random walk process between a spherical catalytic particle of radius $r=R$ and an outer sphere of larger radius $r=L$ where the molecules have the fixed concentrations $\bar{c}_{\rm A}$ and $\bar{c}_{\rm B}$. The simulation of the random walk process is described in Appendix~\ref{app:A}. The system contains a total of $N=522909$ A and B particles with $r<L$. The parameters take the following values:
\be
D=1 \, , \qquad
\kappa = 0.4 \, , \qquad
R = 0.5 \, , \qquad
L=5 \, , \qquad
\bar{c}_{\rm A} = 526.1169 \, , \qquad
\bar{c}_{\rm B} = 473.5751 \, . \label{3D-ex}
\ee
The process is at the crossover between the reaction- and diffusion-limited regimes because ${\rm Da}=2\kappa R/D=0.4$ and $\tilde\Delta =  1 + {\rm Da}\left(1-R/L\right) = 1.36$.  The mean reaction rate is ${\cal J} = 48.55$ and
the zero-time properties are given by
\be
W_0^{(+)} = 652.40  \, , \qquad
W_0^{(-)} = 603.85  \, , \qquad
{\cal D}_0 = 628.13  \, , \qquad
{\cal A}_0 = 0.077329  \, ,
\ee
while the asymptotic properties are
\be
W_{\infty}^{(+)} = 486.13 \, , \qquad
W_{\infty}^{(-)} = 437.58 \, , \qquad
{\cal D}_{\infty} = 461.86 \, , \qquad
{\cal A}_{\infty} = 0.10521 \, .
\ee
These theoretical values compare favorably with the computational results: ${\cal J} \simeq 49.7$ and ${\cal D}_{\infty} \simeq 465$. The stationary profile of the concentration difference $c_{\rm A}-c_{\rm B}$ is depicted in figure~\ref{fig2} and is also in agreement with the theoretical expectation~(\ref{phi}).

The finite-time affinity is shown in figure~\ref{fig3} where the computational results (dots) are compared with theory (solid line) obtained by spatial discretization into $I=1000$ cells.  The affinity is directly measured from simulation data as the slope of $\ln[P(n,t)/P(-n,t)]$ versus the number $n$ of reactive events during the time interval $[0,t]$ (open squares with error bars).  As time increases, the overlap between the probability distributions $P(n,t)$ and $P(-n,t)$ rapidly decreases.  In order to overcome this difficulty, the Gaussian probability distribution
\be
P(n,t) \simeq \frac{1}{\sqrt{2\pi\sigma_t^2}} \, \exp\left[ -\frac{(n-\langle n\rangle_t)^2}{2\sigma_t^2} \right]
\label{Gaussian_fit}
\ee
is fitted to the histogram. No significant deviations between the histogram and the Gaussian distribution have been observed. The affinity is thus estimated as ${\cal A}_t \simeq 2\, \langle n\rangle_t/\sigma_t^2$.
These values are depicted as filled squares in figure~\ref{fig3}, showing agreement between the simulation data and theory.

\vskip 0.3 cm

Microscopic simulations have also been carried out using a hybrid molecular dynamics-multiparticle collision dynamics scheme~\cite{MK99,MK00,K08,GIKW09}. In particular, we use the implementation with reversible catalytic reactions that satisfies detailed balance.~\cite{HSGK18} The catalytic particle resides in a system containing half inert solvent (S) particles and half reactive A and B particles. The roughly spherical catalytic particle is made of catalytic (C) beads connected by stiff harmonic springs. The fluid species interact with the particle beads through repulsive Lannard-Jones potential functions. Reversible reactions, ${\rm C}+{\rm A}\rightleftharpoons{\rm C}+{\rm B}$, take place on the catalytic beads with forward ($p_+$) and reverse ($p_-$) reaction probabilities. The reactive collision rule is designed such that the forward and reverse collisions satisfy the principle of detailed balance. To establish nonequilibrium conditions, the concentrations of A and B particles at a distance $L$ are controlled by relabelling particle species as A with probability $\bar{p}_{\rm A}$ or as B with probability $\bar{p}_{\rm B} = 1- \bar{p}_{\rm A}$ when a reactive particle moves across the boundary at $L$ from the region outside $L$ into the system. The resulting concentrations at $r = L$ are $\bar{c}_{\rm A} = \bar{p}_{\rm A}c_0 $ and $\bar{c}_{\rm B} = \bar{p}_{\rm B}c_0$, where $c_0 = \bar{c}_{\rm A}+\bar{c}_{\rm B}$ is the total concentration of A, B and S particles. This simulates a system where the concentrations outside of the sphere of radius $L$ are prescribed to be $\bar{c}_{\rm A}$ and $\bar{c}_{\rm B}$. The spherical particle is not fixed in space and undergoes Brownian motion. The sphere of radius $L$ is centered on the instantaneous position of the particle. See simulation details in Appendix~\ref{app:B}.

As above, we focus on the case where the diffusion coefficients and the forward and reverse rate constants are equal, $D\equiv D_{\rm A}=D_{\rm B}$ and $\kappa\equiv \kappa_{\pm}$. To determine $\kappa \equiv k^0/(4\pi R^2)$, we consider the irreversible reaction ${\rm C}+{\rm A}\to{\rm C}+{\rm B}$ on the particle with reaction probability $p_+ = 1$. The rate law for this case is $dc_{\rm A}(t)/dt = -k(t)n_{\rm C}c_{\rm A}(t)$, where $n_{\rm C} = 1/V$ is the Janus particle density. The time-dependent rate coefficient $k(t)$ starts at $k(0^+) = k^0$ and decays to the asymptotic value $k = k^0k_D/(k^0 +k_D)$ where $k_D = 4\pi DR$ is the diffusion-limited rate constant~\cite{TK04,HSGK18}. The time-dependent rate coefficient can be found by computing $k(t) = -[dc_{\rm A}(t)/dt]/[n_{\rm C}c_{\rm A}(t)]$ with the system starting from all A particles in the bulk without particle relabelling at the boundary $L$. The values obtained are $k^0\simeq 188.4 \pm 17.6$ and $k \simeq 3.68\pm 0.05$, from which one gets $k_D \simeq 3.75$. From $k_D$, one can estimate the outer edge of boundary layer to be at $R = k_D/(4\pi D)\simeq 5.0$, where $D = 0.0596$ is the fluid species diffusion coefficient. The Damk\"ohler number is equal to ${\rm Da} = 2\kappa R/D \simeq 100$, and $\tilde{\Delta}=80.2$, meaning that the system is in the diffusion-limited regime.

\begin{figure}[h]
\centerline{\scalebox{0.28}{\includegraphics{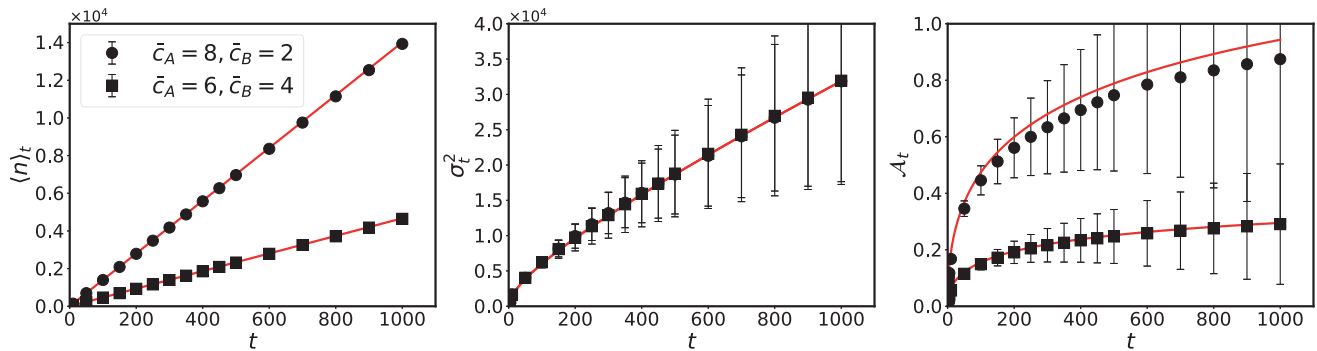}}}
\caption{Spherical catalytic particle: Plots of $\langle n \rangle_t$, $\sigma^2_t$ and the finite-time affinity ${\cal A}_t$ versus time $t$ for two different reservoir conditions indicated in the figure inset. The dots are the simulation data and the solid lines the theoretical results.}
\label{fig4}
\end{figure}

The following two sets of parameter values with different reservoir concentrations have been used to obtain the results:
\be
D=0.0596 \, , \qquad
\kappa = 0.6 \, , \qquad
R = 5.0 \, , \qquad
L=24 \, , \qquad
\bar{c}_{\rm A} = \{8,6\} \, , \qquad
\bar{c}_{\rm B} = \{2,4\} \, . \label{3D-ex-MPC}
\ee
The mean reaction rates are ${\cal J} \simeq \{14.0, 4.7\}$, the zero-time properties are given by
\be
W_0^{(+)} = \{941.7, 937.0\}  \, , \qquad
W_0^{(-)} = \{927.7, 932.4\}  \, , \qquad
{\cal D}_0 = \{934.7, 934.7\}  \, , \qquad
{\cal A}_0 = \{0.015, 0.005\}  \, ,
\ee
while the asymptotic properties are
\be
W_{\infty}^{(+)} = \{18.7, 14.0\}\, , \qquad
W_{\infty}^{(-)} = \{4.7, 9.3\}\, , \qquad
{\cal D}_{\infty} = \{11.7, 11.7\}\, , \qquad
{\cal A}_{\infty} = \{1.4, 0.4\}\, .
\ee
These theoretical values compare favorably with the computational results: ${\cal J} \simeq
\{13.9,4.6\}$ and ${\cal D}_{\infty} \simeq \{13.6, 13.2\}$. The time-dependent $\langle n \rangle_t$, $\sigma_t^2$ and ${\cal{A}}_t$ obtained from theory and simulations are plotted in figure~\ref{fig4} where good agreement is seen.

Comparing figure~\ref{fig3} and figure~\ref{fig4}, we observe that the time-dependent affinity ${\cal A}_t$ starts from an early-time value ${\cal A}_0$ that is much closer to the asymptotic value ${\cal A}_{\infty}$ in the reaction-limited regime than in the diffusion-limited regime.  The reason is that the time-dependent affinity converges faster in the former regime than in the latter regime, as shown in subsection~\ref{S-geom-theory}.


\section{Janus catalytic particle}
\label{J-geom}

\subsection{Theory}

Here, we consider an immobile Janus particle of radius $R$ centered on the origin $r=0$ in the spherical coordinates $(r,\theta,\varphi)$.  The upper hemisphere is catalytic and the lower one is inert, so that the problem has a cylindrical symmetry under the rotations $\varphi\to\varphi+\alpha$ around the axis of the Janus particle.

The stationary problem~(\ref{phi1})-(\ref{phi4}) is given by
\be
\nabla^2\phi =\frac{1}{r}\frac{\partial^2}{\partial r^2}\left(r\, \phi \right) +\frac{\hat{\cal L}}{r^2}\, \phi = 0 \, ,
\ee
with the operator acting as $\hat{\cal L} Y_{lm}= -l(l+1) Y_{lm}$ on the spherical harmonics $Y_{lm}(\theta,\varphi)$.
The boundary conditions read
\be
\left(\frac{\partial\phi}{\partial r}\right)_R = \frac{1}{\ell} (\phi-1)_R H(\cos\theta) \qquad\mbox{and}\qquad (\phi)_L=0 \, ,
\ee
where $H(\xi)$ is Heaviside's function such that $H(\xi)=1$ if $\xi = \cos \theta >0$ and zero otherwise.  The solution of this problem can be expressed as
\be
\phi(r,\theta) = {\rm Da} \sum_{l=0}^{\infty} a_l \, \left[ \left(\frac{R}{r}\right)^{l+1} - \left(\frac{R}{L}\right)^{l+1}\left(\frac{r}{L}\right)^l \right] \, P_l(\cos\theta)
\label{phi-Janus}
\ee
in terms of the coefficients
\be
a_l = \sum_{l'=0}^{\infty} \left({\boldsymbol{\mathsf M}}^{-1}\right)_{ll'} \, \int_0^1 d\xi \, P_{l'}(\xi)
\ee
with
\be
\left({\boldsymbol{\mathsf M}}\right)_{ll'} = \frac{2}{2l+1} \left[ l+1 + l \left(\frac{R}{L}\right)^{2l+1}\right] \delta_{ll'} + {\rm Da} \left[ 1 -\left(\frac{R}{L}\right)^{2l'+1}\right] \int_0^1 d\xi \, P_l(\xi) \, P_{l'}(\xi) \, ,
\label{Mll}
\ee
where Da is the Damk\"ohler number~(\ref{Damkohler}).  The stationary mean concentrations are thus given by equation~(\ref{NESS}).

Here, the effective catalytic surface area~(\ref{Sigma}) is given by $\Sigma = 4\pi R^2 a_0 = 2\pi R^2 \left( 1 -{\rm Da} \, \gamma_J\right)$ with the constant
\be
\gamma_J =\frac{1-2a_0}{\rm Da} = \sum_{l=0}^{\infty} a_l \left[ 1 - \left(\frac{R}{L}\right)^{2l+1}\right] \, \int_0^1 d\xi \, P_l(\xi) \, .
\label{gamma_J}
\ee
In the limit $L\to\infty$, the effective catalytic surface area can be approximated by $\Sigma \simeq 2\pi R^2/(1+ 0.708115\, {\rm Da} )$, as shown in Ref.~\cite{GK18}.

The functions~(\ref{Upsilon}) can here be expressed as
\be
\Upsilon_k(t) = 4\pi \int_R^L dr \sum_{l=0}^{\infty} \frac{1}{2l+1} \, v_{kl}(r,0) \left[ v_{kl}(r,0) - v_{kl}(r,t)\right]  ,
\label{Upsilon-Janus}
\ee
by expanding the functions $f_k(r,\theta,t)$ as
\be
f_k(r,\theta,t) = \frac{1}{r} \sum_{l=0}^{\infty} v_{kl}(r,t) \, P_l(\cos\theta)
\ee
in terms of the solutions of
\be
\partial_tv_{kl}(r,t) = D_k \left[ \partial_r^2 - \frac{l(l+1)}{r^2}\right] \, v_{kl}(r,t)
\label{Janus-vkl}
\ee
with the boundary conditions
\be
\left(\partial_r v_{kl}\right)_R =\frac{1}{R} (v_{kl})_R + \frac{2l+1}{2} \sum_{l'=0}^{\infty} \left(\frac{\kappa_+}{D_{\rm A}} \, v_{{\rm A}l'}+\frac{\kappa_-}{D_{\rm B}} \, v_{{\rm B}l'}\right)_R \int_0^1 d\xi \, P_l(\xi) \, P_{l'}(\xi)
\label{Janus-vkl-bc}
\ee
and $(v_{kl})_L=0$, and starting from the initial conditions
\be
v_{kl}(r,0) = {\rm Da} \, R \, a_l \, \left[ \left(\frac{R}{r}\right)^{l} - \left(\frac{R}{L}\right)^{l}\left(\frac{r}{L}\right)^{l+1} \right]
\label{Janus-vkl-0}
\ee
for $k={\rm A}$ and $k={\rm B}$.

Therefore, the rates are given by equation~(\ref{W}) with
\be
W_{\infty}^{(+)} = 2\pi R^2 \left( 1 -{\rm Da} \, \gamma_J\right) \, \kappa_+ \bar{c}_{\rm A} \, , \qquad\qquad
W_{\infty}^{(-)} = 2\pi R^2 \left( 1 -{\rm Da} \, \gamma_J\right) \, \kappa_- \bar{c}_{\rm B} \, ,
\label{W-inf-Janus}
\ee
and the function (\ref{Psi}) is expressed in terms of the functions~(\ref{Upsilon-Janus}), which behave qualitatively as in the spherical geometry. If $L$ is finite, we have in the long-time limit that
\be
\Upsilon_{\rm A}(\infty) = \Upsilon_{\rm B}(\infty) = \frac{4\pi}{3}\, a_0^2 \, \frac{R^4}{\ell^2} \, L \, \left[1+O(R/L)\right] \qquad\mbox{with}\qquad a_0=\frac{1}{2} \left( 1 -{\rm Da} \, \gamma_J\right) ,
\ee
so that
\be
\Psi(\infty) = \frac{4\pi}{3}\, a_0^2 R^4 L \kappa_+\kappa_- \left(\frac{\bar{c}_{\rm A}}{D_{\rm B}^2} + \frac{\bar{c}_{\rm B}}{D_{\rm A}^2} \right) \left[1+O(R/L)\right] ,
\label{Psi-Janus-t_inf}
\ee
which is proportional to $L$ as in the spherical geometry.  Therefore, the rates~(\ref{W}) converge to their asymptotic value~(\ref{W-inf-Janus}) with corrections of $O(1/t)$, if $L$ remains finite.  However, if $L$ is infinite, the convergence proceeds with corrections of $O(1/\sqrt{t})$, as in the spherical geometry. Consequently, the affinity~(\ref{A-dfn}) converges towards its asymptotic value~(\ref{A-inf}) if $L$ is finite and infinite.

According to equation~(\ref{W-early}), the rates are given at early time by
\bea
W_t^{(+)} &=& 2\pi R^2 \kappa_+  \left[ \bar{c}_{\rm A} - \gamma_J\, \frac{R}{D_{\rm A}} \left(\kappa_+\bar{c}_{\rm A} -  \kappa_-\bar{c}_{\rm B}\right)\right] + O(t) \, ,\\
W_t^{(-)} &=& 2\pi R^2 \kappa_-  \left[ \bar{c}_{\rm B} + \gamma_J\, \frac{R}{D_{\rm B}} \left(\kappa_+\bar{c}_{\rm A} -  \kappa_-\bar{c}_{\rm B}\right)\right] + O(t)\, ,
\eea
so that the early time behavior of the affinity is given by
\be
{\cal A}_t = \ln\frac{\kappa_+  \left[ \bar{c}_{\rm A} - \gamma_J\, \frac{R}{D_{\rm A}} \left(\kappa_+\bar{c}_{\rm A} -  \kappa_-\bar{c}_{\rm B}\right)\right] + O(t)}{\kappa_-  \left[ \bar{c}_{\rm B} + \gamma_J\, \frac{R}{D_{\rm B}} \left(\kappa_+\bar{c}_{\rm A} -  \kappa_-\bar{c}_{\rm B}\right)\right] + O(t)} ,
\ee
which can also be much smaller than the asymptotic affinity~(\ref{A-inf}).


\subsection{Numerical results}

We again suppose that the diffusion coefficients and the rate constants are equal, $D\equiv D_{\rm A}=D_{\rm B}$ and $\kappa\equiv \kappa_{\pm}$.  The finite-time affinity is thus given by equation~(\ref{A-dfn}) with the rates
\bea
W_t^{(+)} &=& 2\pi R^2 \kappa \left(1-{\rm Da}\, \gamma_J\right) \bar{c}_{\rm A} + \ell^2\frac{\kappa^2}{D^2}\left(\bar{c}_{\rm A}+\bar{c}_{\rm B}\right) \frac{\Upsilon(t)}{t} \, ,\\
W_t^{(-)} &=& 2\pi R^2 \kappa \left(1-{\rm Da}\, \gamma_J\right) \bar{c}_{\rm B} + \ell^2\frac{\kappa^2}{D^2}\left(\bar{c}_{\rm A}+\bar{c}_{\rm B}\right) \frac{\Upsilon(t)}{t} \, ,
\eea
where the function $\Upsilon(t)=\Upsilon_{\rm A}(t)=\Upsilon_{\rm B}(t)$ is defined by equation~(\ref{Upsilon-Janus}) and calculated by solving equations~(\ref{Janus-vkl})-(\ref{Janus-vkl-bc}) for $v_l(r,t)\equiv v_{{\rm A}l}(r,t)=v_{{\rm B}l}(r,t)$.  This problem is solved numerically by spatial discretization $v_{l,i}(t) =v_l(r_i,t)$ with $r_i=R + (i-1/2)\Delta r$ where $i=1,2,...,I$ and $\Delta r = (L-R)/I$ with $I=20$ and $l\leq 100$ (see details in Appendix~\ref{app:C}).

\vskip 0.5 cm

Again, the theoretical results are compared with simulations where A and B particles move according to a diffusive random walk process between an immobile Janus particle of radius $r=R$ with a hemispherical catalytic surface and an outer sphere of larger radius $r=L$ where the molecules have the fixed concentrations $\bar{c}_{\rm A}$ and $\bar{c}_{\rm B}$, as described in Appendix~\ref{app:A}. The system contains a total of $N=523164$ A and B particles with $r<L$.  Here, the parameters have the following values:
\be
D=1 \, , \qquad
\kappa = 0.4 \, , \qquad
R = 0.5 \, , \qquad
L=5 \, , \qquad
\bar{c}_{\rm A} = 527.195 \, , \qquad
\bar{c}_{\rm B} = 474.545 \, . \label{Janus-ex}
\ee
The process is at the crossover between the reaction- and diffusion-limited regimes since ${\rm Da} =  2\kappa R/D = 0.4$. For this system, we have that $\gamma_J = 0.51960$. The mean reaction rate is given by ${\cal J} = 26.2 $, the time-zero properties by
\be
W_0^{(+)} = 327.81  \, , \qquad
W_0^{(-)} = 301.60  \, , \qquad
{\cal D}_0= 314.71  \, , \qquad
{\cal A}_0 = 0.083318  \, ,
\ee
and the asymptotic properties by
\be
W_{\infty}^{(+)} = 262.40 \, , \qquad
W_{\infty}^{(-)} = 236.19 \, , \qquad
{\cal D}_{\infty} = 249.30 \, , \qquad
{\cal A}_{\infty} = 0.10521 \, .
\ee
Here also, these theoretical values compare favorably with the computational results: ${\cal J} \simeq 26.5$ and ${\cal D}_{\infty} \simeq 250$. We notice that the diffusivity does not deviate much from its asymptotic value in the present case.

\begin{figure}[h]
\centerline{\scalebox{0.5}{\includegraphics{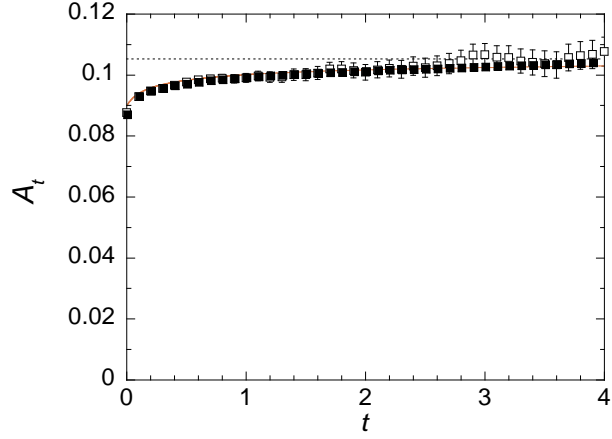}}}
\caption{Janus catalytic particle: Affinity ${\cal A}_t$ versus time $t$ for the simulation with the parameter values~(\ref{Janus-ex}).  The dots are the simulation data and the solid line the theoretical result. The open squares show the affinity directly measured with $\ln[P(n,t)/P(-n,t)]$, while the filled squares show the affinity obtained from the Gaussian fit~(\ref{Gaussian_fit}).}
\label{fig5}
\end{figure}

The finite-time affinity is shown in figure~\ref{fig5} where the computational results (squares) are compared with theory (solid line).  As for the spherical catalytic particle, the affinity is directly measured from simulation data as the slope of $\ln[P(n,t)/P(-n,t)]$ versus the number $n$ of reactive events during the time interval $[0,t]$ (open squares with error bars), as well as by Gaussian fits to the histograms of $n$ values.  No significant deviations between the histogram and the Gaussian distribution have been observed.  The affinity is then estimated as ${\cal A}_t \simeq 2\, \langle n\rangle_t/\sigma_t^2$.  These values are depicted as filled squares in figure~\ref{fig5}, showing here also agreement between the simulation data and theory.

\begin{figure}[h]
\centerline{\scalebox{0.28}{\includegraphics{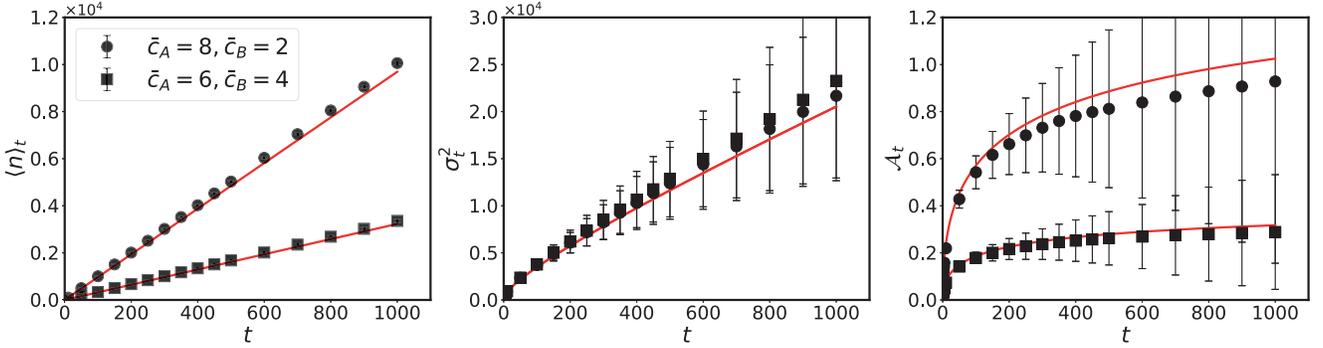}}}
\caption{Janus catalytic particle: The mean number of reactive events $\langle n\rangle_t$, the corresponding variance $\sigma^2_t$, and the affinity~${\cal A}_t$ versus time $t$ for the microscopic simulation with the multiparticle collision method.  The dots are the simulation data and the solid lines the theoretical results.  The results are the average of $40$ realizations and the error is given by the standard deviation.}
\label{fig6}
\end{figure}

Microscopic simulations as described in Appendix~\ref{app:B} have also been carried out for a Janus particle made from catalytic and noncatalytic beads connected by stiff harmonic springs. We have chosen the interaction potentials of the A and B particles with the beads of the Janus particle to be equal so that the diffusiophoretic mechanism that leads to self propulsion does not operate. The Janus particle does, however, execute Brownian motion as is the case for the spherical particle. Again, we investigate the time-dependent reaction rates, diffusivities and affinity for two parameter sets with different reservoir concentrations:
\be
D=0.0596 \, , \qquad
\kappa = 0.6 \, , \qquad
R = 5.0 \, , \qquad
L=24 \, , \qquad
\bar{c}_{\rm A} = \{8,6\} \, , \qquad
\bar{c}_{\rm B} = \{2,4\} \, . \label{3D-Janus-MPC}
\ee
Here, the process evolves in the diffusion-limited regime because the Damk\"ohler number has the value ${\rm Da}=2\kappa R/D\simeq 100$.
The mean reaction rates are ${\cal J} \simeq \{9.7,3.23\}$, the zero-time properties are given by
\be
W_0^{(+)} = \{475.8, 472.6\}  \, , \qquad
W_0^{(-)} = \{466.1, 469.4\}  \, , \qquad
{\cal D}_0 = \{471.0, 471.0\}  \, , \qquad
{\cal A}_0 = \{0.021,0.0068\}  \, ,
\ee
while the asymptotic properties are
\be
W_{\infty}^{(+)} = \{12.9, 9.7\}\, , \qquad
W_{\infty}^{(-)} = \{3.23, 6.47\}\, , \qquad
{\cal D}_{\infty} = \{8.09, 8.09\}\, , \qquad
{\cal A}_{\infty} = \{1.38,0.405\}\, .
\ee
These theoretical values compare favorably with the simulation results: ${\cal J} \simeq
\{10.1,3.35\}$ and ${\cal D}_{\infty} \simeq \{9.3,10.6\}$. The time-dependent $\langle n \rangle_t$, $\sigma_t^2$ and ${\cal{A}}_t$ obtained from theory and simulations are plotted in figure~\ref{fig6} where good agreement is seen.

Here also, the comparison between figure~\ref{fig5} and figure~\ref{fig6} shows that the time-dependent affinity ${\cal A}_t$ remains closer to its asymptotic value ${\cal A}_{\infty}$ in the reaction-limited regime than in the diffusion-limited regime, confirming the general behavior already observed in the spherical geometry.

\section{Conclusion and perspectives}
\label{conclusion}

In this paper, the finite-time fluctuation theorem of Ref.~\cite{TDFT} has been investigated for the diffusion-influenced surface reaction ${\rm A}\rightleftharpoons{\rm B}$ on spherical and Janus catalytic particles.  The finite-time rates of the forward and reverse reactions ${\rm A}\rightleftharpoons{\rm B}$ have been analytically calculated in both geometries by solving diffusion equations with the special boundary conditions obtained in Ref.~\cite{TDFT}.  These rates provide the time-dependent thermodynamic force or affinity driving the process away from equilibrium.  This affinity converges towards its asymptotic value determined by the concentrations of the reacting species at the reservoir, as predicted by infinite-time fluctuation theorems.

The results show that the affinity may take a much lower value at early time than its expected asymptotic value.  The reason is that the affinity reaches its asymptotic value beyond the diffusion time characteristic of the reaction taking place on the catalytic particle.   In the reaction-limited regime, this diffusive time is short, so that the affinity rapidly converges towards its asymptotic value.  However, the diffusion time may be significantly longer in the diffusion-limited regime.  For micrometric catalytic particles and small diffusing molecules, the crossover time is of the order of miliseconds, which is short relative to macroscopic measurement times, hence justifying the use of the long-time limit.

Theoretical results for the spherical catalytic and Janus particles were compared with numerical simulations using two different methods: a random walk algorithm and an algorithm based on multiparticle collision dynamics. These systems were studied for diffusion-limited catalytic reactions as well as for reactions that lie in crossover regime between reaction- and diffusion-limited kinetics. In the diffusion-limited regime the affinity takes an early-time value that is significantly smaller than its asymptotic value. In all cases theoretical and simulation results are in agreement.

In the three-dimensional geometries of the spherical and Janus catalytic particles, the crossover time does not depend on the distance between the catalytic surface and the reservoir where the concentrations of reacting species are fixed.  This is no longer the case in the one-dimensional planar geometry, as we shall report in a future publication.

\section*{Acknowledgments}

Financial support from the International Solvay Institutes for Physics and Chemistry, the Universit\'e libre de Bruxelles (ULB), the Fonds de la Recherche Scientifique~-~FNRS under the Grant PDR~T.0094.16 for the project ``SYMSTATPHYS", the Belgian Federal Government under the Interuniversity Attraction Pole project P7/18 ``DYGEST", and the Natural Sciences and Engineering Research Council of Canada is acknowledged.


\appendix
\section{Random walk simulation method and parameters}\label{app:A}

The system contains $i=1,2,...,N$ particles moving in a cubic box of size
$\mathscr{L}>L$ according to the Langevin stochastic differential equations
$d{\bf r}_i/dt = {\bf v}_i(t)$ where the velocities are given by Gaussian white
noises satisfying $\langle{\bf v}_i(t)\rangle = 0$ and
$\langle{\bf v}_i(t)\otimes{\bf v}_j(t')\rangle = 2\,D\,\delta_{ij}\,\delta(t-t')\,
{\boldsymbol{\mathsf 1}}$, expressed in terms of the diffusion coefficient
$D$ and $3\times 3$ identity matrix ${\boldsymbol{\mathsf 1}}$.
These equations are solved numerically by discretization into time steps $\Delta t=0.001$.
Initially, the cubic box is uniformly filled with $10^6$ particles, so that the overall particle density is equal to $c_0=10^6/\mathscr{L}^3$.  As the particles move into a sphere of radius $L$ centered at the origin inside the box, they acquire a color A or B with the probabilities $\bar{P}_{\rm A}=\bar{c}_{\rm A}/c_0$ or $\bar{P}_{\rm B}=\bar{c}_{\rm B}/c_0=1-\bar{P}_{\rm A}$. Simultaneously, particles crossing the surface $r=L$ are specularly reflected by the surface, so that the number of particles inside the sphere of radius $L$ is constant and equal to $N=c_0\times 4\pi L^3/3$. This determines the boundary values of the concentrations at $r=L$.

The color A or B of the moving particles changes upon their collision with the catalytic surface.
The catalyst occupies a spherical domain of radius $R<L$ centered at the origin.
The surface reaction is simulated according to the algorithm of Ref.~\cite{SSOH08}.
Whenever the particle displacement ${\bf r}_i(t+\Delta t)-{\bf r}_i(t) = {\bf \Delta}_i(t)$
crosses the catalytic surface the moving particle changes its color with the probabilities $P_{\pm}= \kappa_{\pm}\sqrt{\pi\Delta t/D}$ and simultaneously its trajectory
is specularly reflected by the surface.
There is a chance that a particle hits the catalyst even if ${\bf r}_i(t+\Delta t)$ is located
outside the spherical domain. That is during the time interval $[t, t+\Delta t]$ the particle
might have crossed the catalyst surface twice. If ${\bf \Delta}_i(t)$ intersects
the surface at two locations, the particle trajectory is reflected specularly at the first
intersection point.

In the reported simulations, the following values are taken for the parameters: $\mathscr{L}=10$, $L=5$, $R=0.5$, $D=1$, $\bar{P}_{\rm A}=10/19$, $\bar{P}_{\rm B}=9/19$, and $c_0=\bar{c}_{\rm A}+\bar{c}_{\rm B}=10^3$.  Time series with $10^6$ data points have been computed for the number of reactive events during time interval $10\times \Delta t$ of the simulation.  In Eqs.~(\ref{3D-ex}) and~(\ref{Janus-ex}), the given values of the concentrations have been determined from the mean numbers of particles of both species in the layer $L-\Delta r < r < L$ next to the reservoir [with $\Delta r = (L-R)/50=0.09$].

\section{Microscopic simulation method and parameters}\label{app:B}

Here, we describe the hybrid molecular dynamics-multiparticle collision dynamics scheme and give the parameter values used in the simulations.

The Janus motor of radius $R = R_J + \sigma = 5\,\sigma$ is placed in a cubic simulation box of linear length $\mathscr{L}=50\,\sigma$ containing $N = N_{\rm A}+N_{\rm B}+N_{\rm S} = 2488439$ fluid particles with $N_{\rm A}+N_{\rm B} = N_{\rm S}$. The average densities of the fluid and reactive particles are $N/\mathscr{L}^3 \simeq 20$ and $c_0 = (N_{\rm A}+N_{\rm B})/\mathscr{L}^3 \simeq 10$, respectively. The construction of the Janus motor was described earlier~\cite{HSGK18}. It is made of $2681$ beads randomly distributed in a sphere of radius $R_J = 4\,\sigma$, where two beads within a distance $2\,\sigma$ are linked by a stiff spring with spring constant $k_s = 50\,k_{\rm B}T/\sigma^2$, where $k_{\rm B}T$ is the thermal energy. The interaction between a motor bead and a fluid particle is given by a repulsive Lennard-Jones potential with interaction strength~$\epsilon_{\alpha}$, $U_{\alpha}(r) = 4\,\epsilon_{\alpha} [(\sigma/r)^{12} - (\sigma/r)^6 + 0.25]$, which vanishes when $r> 2^{1/6}\sigma$. The interaction strengths are chosen as $\epsilon_{\rm A} = \epsilon_{\rm B} = 1.0$.

The nonequilibrium steady states discussed in the main text are established by considering a spherical region with radius $r = L = 24\:\sigma$ centered on the Janus motor. The region outside of $r=L$ is modeled as a reservoir with prescribed concentrations of A and B species, which can be controlled by changing the species type to A or B with probabilities $\bar{p}_{\rm A}$ and $\bar{p}_{\rm B}$ with $\bar{c}_{\rm A} = \bar{p}_{\rm A} c_0$ and $\bar{c}_{\rm B} = \bar{p}_{\rm B} c_0$. In the simulations we consider (a) $\bar{p}_{\rm A} = 0.8$ and $\bar{p}_{\rm B} = 0.2$ and (b) $\bar{p}_{\rm A} = 0.6$ and $\bar{p}_{\rm B} = 0.4$. Note that there is no change of species for inert S particles at the boundary $r=L$.

The dynamics of fluid particles is described by multiparticle collision dynamics comprising streaming and collision steps at discrete time intervals $\tau = 0.1\,t_0$. The collisions are carried out by first sorting the particles into a grid of cubic cells with linear size $\sigma$ and the postcollision velocities of particle $i$ in a cell $\xi$ are given by $\mathbf{v}'_i = \mathbf{V}_{\xi} + \hat{\cal R} ( \mathbf{v}_i - \mathbf{V}_{\xi})$, where $\mathbf{V}_{\xi}$ is the center of mass velocity of particles in cell $\xi$ and $\hat{\cal R}$ is a rotation operator about a random axis by an angle of $120^{\circ}$. Between two consecutive collisions, the system evolves by Newton's equation of motion with forces determined from the total potential energy of the system using a time step of $\delta r = 0.005\,t_0$. The common diffusion coefficient of fluid particles is found to be $D = 0.0596$. Simulation results are reported in dimensionless units where mass is in units of $m$, length in units of $\sigma$, energies in units of $k_{\rm B}T$ and time in units of $t_0 = \sqrt{m \sigma^2 / k_{\rm B}T}$.

\section{Discretization of the diffusion equations~(\ref{Janus-vkl}) and the boundary conditions~(\ref{Janus-vkl-bc})}\label{app:C}

The numerical method for solving equations~(\ref{Janus-vkl}) and~(\ref{Janus-vkl-bc}) for $v_{kl}(r,t)$ is as follows. We suppose the diffusion coefficients and the rate constants are equal, $D\equiv D_{\rm A} = D_{\rm B}$ and $\kappa \equiv \kappa_{\pm}$ and, therefore, one obtains $v_l(r,t) \equiv v_{{\rm A}l}(r,t)=v_{{\rm B}l}(r,t)$. The discretized diffusion equation for $v_l(r,t)$ is
\begin{equation}
v_l(r,t+\Delta t) = \bigg[ 1 - 2 \frac{D\Delta t}{\Delta r^2}  - \frac{l(l+1)}{r^2}D\Delta t  \bigg]   v_l(r, t) + \frac{D\Delta t}{\Delta r^2} \bigg[ v_l(r+\Delta r, t) + v_l(r-\Delta r, t) \bigg] ,
\end{equation}
subject to the boundary condition at $r =L$, $v_l(r = L,t)=0$, and the boundary condition at $r = R$,
\begin{equation}
v_l(R,t) =\sum_{l'=0}^{\infty} \Big(\tilde{\boldsymbol{\mathsf M}}^{-1}\Big)_{ll'}\, v_{l'}(R+\Delta r,t) \, ,
\end{equation}
where
\begin{equation}
\big(\tilde{\boldsymbol{\mathsf M}}\big)_{ll'} = \bigg( 1+\frac{\Delta r}{R}\bigg)\, \delta_{ll'} + (2l+1)\, \frac{\kappa\Delta r}{D}  \int_0^1 P_l(\xi) P_{l'}(\xi)\, d\xi
\end{equation}
with the Kronecker delta symbol $\delta_{ll'}$.


\end{document}